\documentclass[onecolumn,final]{elsart3p}
\usepackage{amsmath, amssymb, amsfonts, graphicx, subfigure}
\journal{Physica D: Nonlinear Phenomena}

\def\d{\bar{d_0}}
\def\Schrod{Schr$\ddot{\mbox{o}}$dinger }

\begin{document}

\begin{frontmatter}

\title{New Bisoltion Solutions in Dispersion Managed Systems}

\author{M.~Shkarayev\corauthref{cor}},
\corauth[cor]{Corresponding author.}
\ead{max@math.arizona.edu}
\author{M.G.~Stepanov}

\address{Department of Mathematics, The University of Arizona, 617 N.
Santa Rita Ave., Tucson, AZ 85721, USA}

\begin{abstract}
  In this paper we propose a method which provides a full description of solitary
  wave solutions of the \Schrod equation with periodically varying
  dispersion. This method is based on analysis and polynomial deformation of the spectrum of an iterative map. Using this method we discover a new family of antisymmetric bisoliton solutions.
  In addition to the fact that these solutions are of interest for nonlinear fiber optics and the theory of nonlinear \Schrod equations with periodic coefficients, they have potential applications for increasing of bit-rate in high speed optical fiber communications.
\end{abstract}

\end{frontmatter}

\section{Introduction}

Chromatic dispersion of optical fibers is one of the major factors
limiting capacity of fiber communication systems.  It results in
broadening of optical pulses due to difference in velocities of
different spectral components. The dispersion management
technique~\cite{Linn-Kogelnik-80} was proposed as a method to solve the
problem of dispersive pulse broadening. An optical fiber link with
dispersion management is composed of periodically arranged fiber spans
with alternating signs of dispersion. The fiber spans are chosen to make
the cumulative effect of dispersion small or even zero. Full
compensation of the dispersive effects is achieved by making the mean
value of the dispersion to be zero. In this case as a pulse propagates
over one period it will be spread and then contracted to its original
shape.

Increasing bit-rates inevitably leads to manifestation of nonlinear
properties of optical fiber. As bit rates increase, the temporal width
of an optical pulse (bit carrier) $\tau_0$ decreases. On the other hand
the energy of the pulse must remain above some critical level $
{\mathcal E}\geq {\mathcal E}_{\scriptsize{\mbox{cr}}}$ to provide
minimal detection bit-error at the end of the line, and therefore the
power of the signal ($P \sim {\mathcal E}/\tau_0 $) increases as the
pulse narrows $P \geq {\mathcal E}_{\scriptsize{\mbox{cr}}}/\tau_0$. It
is a well known, experimentally verified fact that the index of
refraction of the fiber core grows linearly with the power of the
signal, $n = n_0 + \alpha P$ (where $n_0$ is linear index of refraction
and $\alpha$ is a coefficient of Kerr nonlinearity). With the high
enough power the nonlinear response of the medium becomes important.
Since dispersion management is a purely linear approach to compensate
pulse broadening, validity of this approach becomes less obvious in the
presence of nonlinearity. It was shown that in presence of nonlinearity
the compensation is still possible for a special class of pulses known
as the dispersion managed (DM) solitons~\cite{{KFD-95}}. Detailed
information about the DM solitons and their applications can be found
in~\cite{MolGor_06}.

Soon after the concept of DM solitons was introduced it became clear
that both analytic representation and accurate numerical description of
the DM soliton shape are challenging problems. The first problem was
addressed in many papers, where approximations of the central part of
the DM soliton and asymptotic of oscillatory tails was considered
(see~\cite{KuHas97,TurMez98,MikNov01}). However the general solution of
the this problem remains unsolved. The problem of an accurate numerical
description of the DM soliton shape is also stimulated considerable
effort, which was solved by in~\cite{LushOL00,LushJETP00,Lush01} using
slow dynamics model of DM solitons proposed in~\cite{GT_96}. Later this
approach used to address behavior of DM solitons for the systems with
both dispersion and nonlinearity compensation~\cite{GL02}.

The dispersion compensation while resolving the physical problem with
the pulse broadening, poses an obstacle for the numerical computation of
such solutions. Following the standard approach the description of
soliton consists of launching a signal and waiting for all the
continuous radiation to escape. For the problem with dispersion
management this process is extremely slow and ineffective on the tails
of the soliton. Authors of~\cite{LushOL00} proposed solving of the
averaged equation for the analysis of solutions, which for the solitary
waves reduces from an integro-differential equation to a nonlinear
integral equation, solvable by a nonlinear iterative procedure. This
allowed for a fast way of achieving precision exceeding any necessary
requirements for the practical applications.

Unlike the conventional NLS soliton, the dispersion managed system
supports propagation of bound pairs of solutions, or bisolitons.
Existence of bisoliton solutions was shown in the numerical
investigations of~\cite{Maruta_02} and later confirmed experimentally
in~\cite{Mitschke_05}. It should be noted that this experimental work
considered antisymmetric bisolitons and therefore we limit our
investigation to that type of bisolitons. These solutions are
interesting for the theory of nonlinear partial differential equations,
and represent a new phenomena in nonlinear fiber optics. They illustrate
a new type of solutions of nonlinear \Schrod equation where dispersion
coefficient is replaced with a periodic function. The potential
applications for the optical communication systems were discussed
in~\cite{Mitschke_05}. Bisolitons can be used to introduce a
three-letter alphabet (``empty slot'', ``DM soliton'' and
``bisoliton''), replacing a conventional binary alphabet. This could
increase a system capacity without a change to the physical parameters
of the system.

The shape as well as other important characteristics of bisolitons were
found in~\cite{GIMLSS_07} using a modification of the numerical method
proposed in~\cite{LushOL00, LushJETP00, Lush01}. The equation describing
averaged dynamics of the solution is reduced to have a single universal
parameter, which incorporates all of the physical and geometrical
characteristics of the optical system~\cite{GIMLSS_07}. Analysis of this
equation showed the existence of two branches of bisolitons determined
by the universal parameter, with a critical bifurcation point connecting
the two branches.

Investigation of DM bisolitons was based on the iteration of nonlinear
map. The analysis of the newly discovered branch of solutions was
complicated by the lack of convergence of the iterative procedure.
Careful analysis of the map showed the existence of eigenvalues outside
of the unit circle, explaining the existence of the growing modes of the
map. In this work we propose to utilize the method of polynomial
spectral transformations to construct a contractive map, preserving the
original fixed points. The application of this method allowed for the
discovery of the previously unknown branch of bisoliton solutions. These
bisolitons present new solutions to nonlinear \Schrod equation with
periodic dispersion. A recent paper~\cite{Mitschke_07} indicates a new
experimental evidence of existence of this branch of bisoliton
solutions. The new found bisoliton solutions would allow to introduce a
4-letter alphabet, further increasing the capacity of existing lines.

\section{Basic equations}

A well established model for propagation of an electromagnetic pulse
through a nonlinear, dispersive medium is the nonlinear \Schrod equation
\begin{equation}
  {\rm i} u_z + d(z) u_{tt} + \gamma |u|^2 u = 0
  \label{NLS}
\end{equation}
where the scalar function $u(z, t)$ is the envelope of the signal, $z$
is the distance along the fiber, $t$ is retarded time. Without loss of
generality we consider a fiber link with constant nonlinearity given by
$\gamma$. Function $d(z)$ is the dispersion as a function of distance
along the fiber. For a dispersion managed system this is a piecewise
constant, periodic function which can be written in the form $d(z) = d_0
+ d_1$ if $0 < z < L/2$, and $d(z) = d_0 - d_1$ if $L/2 < z < L$, where
$L$ is a dispersion map period, and $d_1 \gg d_0$.

A fiber link with dispersion management has a characteristic time scale
$\tau_{\scriptsize{\mbox{dm}}}\equiv (Ld_1/2)^{1/2}$. The physical
meaning of this time is the width of a pulse such that the pulse will
approximately double in width as it propagates over half of the period.
The characteristic ``residual dispersion length''
$z_{\scriptsize{\mbox{rd}}} \equiv \tau_{\scriptsize{\mbox{dm}}}^2/d_{0}
= L d_1 / 2d_0 $ is the distance when a pulse with width
$\tau_{\scriptsize{\mbox{dm}}}$ propagating over a fiber with dispersion
$d_0$ will approximately double in width.
Rescaling the variables $t = \tau_{\scriptsize{\mbox{dm}}} \tilde{t}$,
$u = 2 \pi \sqrt{P} \tilde{u}$ results in the equation
\begin{equation}
  {\rm i} \tilde{u}_{z} + \frac{2}{L}
  \frac{d(z) - d_0}{d_1} \tilde{u}_{\tilde{t}\tilde{t}}+
  \frac{1}{z_{\scriptsize{\mbox{rd}}}} \tilde{u}_{\tilde{t} \tilde{t}} +
  (2 \pi)^2
  \frac{1}{z_{\scriptsize{\mbox{nl}}}}|\tilde{u}|^{2}\tilde{u} = 0.
\label{dless}
\end{equation}
$z_{\scriptsize{\mbox{nl}}}\equiv 1/(\gamma P)$ is characteristic
``nonlinear length'' on which the change of the phase of the pulse due
to fiber nonlinearity will be of the order $1$.

To take advantage of dispersion management in the presence of
nonlinearity, the compensation must take place on the distances where
nonlinearity effects are small and negligible compared to the local
value of dispersion. Therefore the dispersion management works in the
linear regime, and two well separated scales of pulse propagation can be
distinguished. On the first scale local dispersion will dominate over
the weak effects the residual dispersion and the nonlinearity. There the
pulse is rapidly broadened and brought back to its original width as a
result of the alternating signs of local dispersion. The second scale is
where the effects of nonlinearity and residual dispersion accumulate and
become important. Thus, the spectrum of the signal can be considered in
the form
\begin{equation}
  F[\tilde{u}] = q(\Omega, z)\exp \left( -{\rm i}\Omega^2
    /(L/2)
    \int_{L/4}^{z} \mbox{d}\xi \, \frac{d(\xi) - d_0}{d_1} \right),
  \label{envelope_phase}
\end{equation}
with the Fourier transform $F$ defined as $F[f(t)] = \int \mbox{d}t
\, \exp({\mbox i} \omega t) f(t)$. Here the exponent is the rapidly
changing phase due to the large value of local dispersion and the
amplitude $q$ is varying slowly with $z$ due to residual
dispersion and the nonlinear effects. Substituting this form of
$\tilde{u}$ into the equation~(\ref{dless}) and taking an average over a
map period we obtain Gabitov-Turitsyn equation describing evolution of
$q$ \cite{GT_96}
\begin{equation}
   \begin{array}{l}
  i q_{z} - \displaystyle \frac{1}{z_{\scriptsize{\mbox{rd}}}}\Omega^2 q
+ \frac{1}{z_{\scriptsize{\mbox{nl}}}} R(q,\Omega) = 0,\\
R(q,\Omega)=
\displaystyle \int {\frac{\sin(\Delta /2)}{\Delta /2}q(\omega_1) q(\Omega_2)q^*(\Omega_3) \delta(\Omega_1 + \Omega_2 - \Omega_3 - \Omega) \mbox{d}
\Omega_1 \mbox{d} \Omega_2 \mbox{d} \Omega_3 },
   \end{array}
\label{Slow}
\end{equation}
where $\Delta \equiv \Omega_1^2+\Omega_2^2 - \Omega_3^2 - \Omega^2$, and
$\delta(\cdot)$ is Dirac's delta function.

We are interested in computation of antisymmetric bisolitons which are
solitary wave solutions of this equation. For such a solution the
amplitude profile is independent of the distance along the fiber, and
can be written as $q(\Omega,\tilde{z})=\varphi(\Omega)e^{i\lambda z}$,
where $\lambda$ is a wave number of the solitary wave. With $\lambda =
1/z_{\scriptsize{\mbox{nl}}}$ the function $\varphi$ is governed by the
following equation
\begin{equation}
  \varphi + \d \Omega^2 \varphi = R(\varphi,\Omega), \mbox{ where
}~~ \d \equiv \frac{z_{\scriptsize{\mbox{nl}}}}{z_{\scriptsize{\mbox{rd}}}} = \frac{d_0}{( \gamma P )
(d_1 L/2) } \label{int_eq}
\end{equation}
Thus far we have reduced this three parameter partial differential
equation to an integral equation with a single free parameter (an
alternative way to do that is presented in
Appendix~A).

\section{Solving the integral equation}

\begin{figure}
  \begin{picture}(370,205)(0,0)
    \put(10,10){\includegraphics[width=0.98\textwidth]{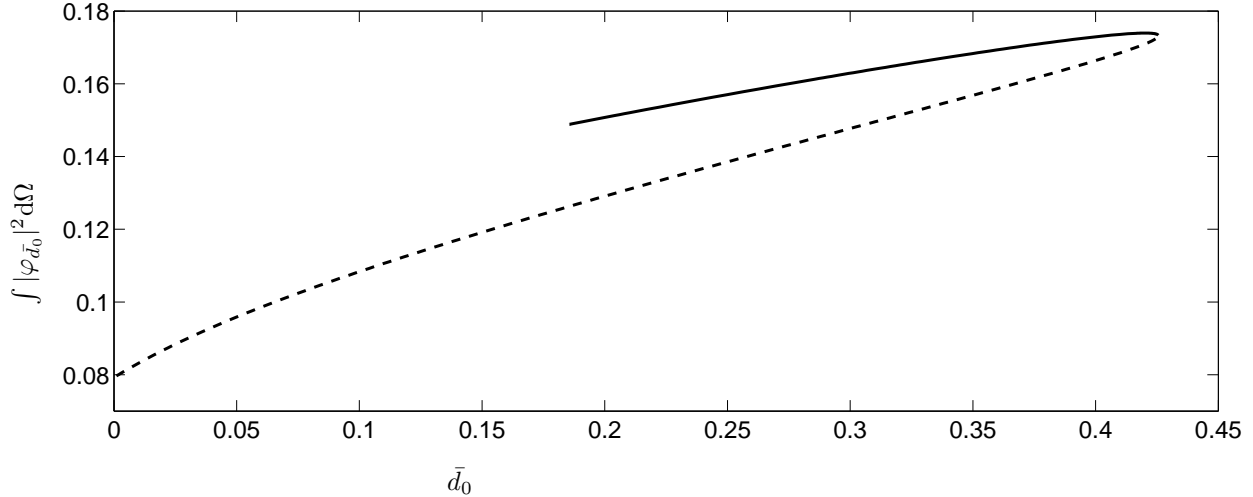}}
    \put(165,0){$\d$}
    \put(0,70){\rotatebox{90}{$\int |\varphi_{\d}|^2 \mbox{d}\Omega$}}
  \end{picture}
  \caption{Energy of the two branches of bisolitons as a function of $\d$. The solid line corresponds to the solutions with higher energy (upper
   branch) and the dashed line corresponds to solutions with lower energy (lower branch)}
  \label{ul_energy}
\end{figure}

In this paper we study the family of antisymmetric bisolitons at
different positive values of parameter $\d$. In order to solve
equation~(\ref{int_eq}) we use the following iterative procedure
\cite{Lushnikov01}
\begin{equation}
 \begin{array}{l}
  \varphi_{n+1} =N_{\d}(\varphi_{n}),~~~ \displaystyle{N_{\d}(\varphi) \equiv
  \mathbb{P}_{\scriptsize{\mbox{odd}}}\left(Q^{3/4}\frac{R(\varphi,\Omega)}{1+ \d
  \Omega^2}\right)}\\ Q \equiv\displaystyle{ \frac{\int|\varphi|^2 \mbox{d}\Omega}{ \int|R(\varphi,\Omega)/\left(1+\d\Omega^2 \right)|^2
  \mbox{d}\Omega}}
 \end{array} \label{scheme}
\end{equation}
where $Q$ is a Petviashvilli factor~\cite{Petviashvili76,
Petviashvili92} which is
introduced to avoid the convergence to the trivial solution $\varphi =
0$ by making that solution unstable. The operator
$\mathbb{P}_{\scriptsize{\mbox{odd}}}\left(f(\Omega)\right) \equiv
\left(f(\Omega)-f(-\Omega)\right)/2$
is a projection onto odd functions, used here to project the iterated
functions onto the space of antisymmetric functions. We solve the
equation~(\ref{scheme}) numerically, discretizing $\varphi$ on a grid of
$M$ points. The iterations are stopped when $\varphi_{n}$ satisfies the
condition
\begin{eqnarray}
  \begin{array}{l} \| \varphi_n + \d\Omega^2 \varphi_n -
    R(\varphi_n,\Omega) \|_{L_2}<10^{-8} \end{array}
  \label{res}
\end{eqnarray}
A Fourier transform of a sum of two Gaussian pulses, shifted apart and
taken with opposite signs is used as a seed for the iteration procedure
with $\d = 0.2$. The value of $\d$ is decreased incrementally, with a
previous solution used as a seed for the next value of $\d$, until the
parameter reaches zero. The same procedure was performed in the
increasing direction from $\d=0.2$. We found two branches of bisoliton
solutions. Figure~\ref{ul_energy} shows the energy of the solutions as a
function of $\d$. The solutions bifurcate near
$\d_{\scriptsize{\mbox{bf}}} \approx 0.4256$. We observed no convergence
for the values of $\d > \d_{\scriptsize{\mbox{bf}}}$.

Direct application of the scheme in equation~(\ref{scheme}) demonstrates
convergence only for the lower branch of solutions, and does not allow
to find the upper branch of solutions. In order to find the solutions
from the upper branch we study the spectrum of the $N$ and modify the
map.

\section{Convergence of the map $N$}\label{spec}

In order to understand the convergence of the iterative procedure
described in equation~(\ref{scheme}) we analyze the spectrum of $N_{\d}$
near a fixed point. In the neighborhood of the fixed point
$\varphi_{\scriptsize{\mbox{fp}}}$ the map $N_{\d}$ is approximated by
the linear operator $K$:
\begin{equation}
   N(\varphi_{\scriptsize{\mbox{fp}}} + \Psi) =
     \varphi_{\scriptsize{\mbox{fp}}} + K(\Psi) + \mathcal{O}(\Psi^2).
\end{equation}
Here the operator $K$ lacks the complex structure and should be
viewed as a linear operator over the field of real numbers acting on a
real vector space consisting of functions $\mbox{Re} \Psi$ and
$\mbox{Im} \Psi$. Let the finite set $\{\mu_i\}$, $\{ | \phi_i \rangle
\} $, and $\{ \langle \phi_i | \} $ correspond to eigenvalues, right
eigenvectors and left eigenvectors of $\hat{K}$, the discretized version
of $K$. Right away we see that there is an eigenvalue 1 corresponding to
a constant phase shift of any fixed point. This eigenvector does not
effect the convergence since the equation~(\ref{res}) is unaffected by
the constant phase shift. Therefore for convenience we introduce sets
$\{ \tilde{\mu}_i \} =\{\mu_i\} \backslash 1$ and $\{ | \tilde{\phi}_i
\rangle \} = \{ | \phi_i \rangle \} \backslash |\phi_{\mu=1} \rangle$,
$\{ \langle \tilde{\phi}_i | \} = \{ \langle \phi_i | \} \backslash
\langle\phi_{\mu=1} |$ where we have removed the above eigenvector and
the corresponding eigenvalue. Furthermore, assume that if $i<j$ than
$|\tilde{\mu}_i|\ge |\tilde{\mu}_j|$, so that $\tilde{\mu}_1$ has
maximal magnitude. The following iterative procedure is used to obtain
the $N^{\scriptsize{\mbox{th}}}$ eigenfunction
\begin{equation}
  |\tilde{\phi}_{N}\rangle= \lim_{n \rightarrow \infty} \Psi_n \mbox{,
  where } \Psi_{n+1}=\frac{\hat{K}(\Psi_n) - \sum_{j=1}^{N-1}
  \langle \tilde{\phi}_j|\hat{K}(\Psi_n) \rangle
  |\tilde{\phi}_j\rangle}{\int |\Psi_n|^2 \mbox{d} \Omega}
  \label{eigenN}
\end{equation}
where $\langle \tilde{\phi}_j|$ is a dual vector of $|\tilde{\phi}_j\rangle$,
satisfying $\langle \tilde{\phi}_j|\tilde{\phi}_k \rangle =\delta_{jk}$ (here
$\delta_{jk}$ is a Kronecker symbol). We assume here that initial
function $\Psi_0$ has nonzero projection onto $N^{th}$ eigenfunction. At
each step of this procedure the largest $N-1$ modes are projected out of
$\Psi_n$ and in the limit as the number of iterations $n$ goes to
infinity the next largest mode is the only one that survives. The
contribution from the $M^{\scriptsize{\mbox{th}}}$ vector ($M>N$) dies
out as $(\tilde{\mu}_M/\tilde{\mu}_N)^n$.

In order to perform the procedure described in equation~(\ref{eigenN})
we use the explicit expression for the operator $\hat{K}$
\begin{equation}
   \begin{array}{l}
     \hat{K}(\Psi) = \displaystyle{ 1.5 \varphi_{\scriptsize{\mbox{fp}}} \int \mbox{Re} \left [ \varphi_{\scriptsize{\mbox{fp}}}^{*}
     \left(\Psi-L(\Psi)\right) \right] \mbox{d}\Omega \bigg/ \int |\varphi_{\scriptsize{\mbox{fp}}}|^2 \mbox{d}\Omega + L(\Psi)}\\
     \mbox{and } L(\Psi) \equiv \displaystyle{\int \frac{\sin (\Delta/2)}{\Delta/2}
     \left( \varphi_{\scriptsize{\mbox{fp}}}(\Omega_1)\varphi_{\scriptsize{\mbox{fp}}}(\Omega_2)\Psi^*(\Omega_3) +\right.}\\
     \displaystyle{\left. +2\varphi_{\scriptsize{\mbox{fp}}}(\Omega_1)\Psi(\Omega_2)\varphi_{\scriptsize{\mbox{fp}}}^*(\Omega_3)\right)
     \frac{\delta(\Omega_1+\Omega_2-\Omega_3-\Omega)}{1+\d\Omega^2}{\bf \mbox{d} \bar{\Omega}}}
   \end{array}
\end{equation}
here ${\bf \mbox{d} \bar{\Omega}}=\mbox{d}\Omega_1 \mbox{d}\Omega_2
\mbox{d}\Omega_3$. The operator $\hat{K}$ is not self-adjoint,
therefore, its left and right eigenvectors are different. We obtain the
the left eigenvectors of the operator $\hat{K}$ by studying the right
eigenvectors of the operator $\hat{K}^{\dagger}$
\begin{equation}
   \begin{array}{l}
    \lefteqn{\hat{K}^{\dagger}(\xi) = L^{\dagger}\left(\zeta\right) - \zeta + L^{\dagger}(\xi)}\\
    \zeta \equiv 1.5 \varphi_{\scriptsize{\mbox{fp}}} \int \mbox{Re} \left[ \varphi_{\scriptsize{\mbox{fp}}}^{*} \xi \right] \mbox{d}\Omega \bigg/
    \int |\varphi_{\scriptsize{\mbox{fp}}}|^2 \mbox{d}\Omega\\
    L^{\dagger}(\xi)=\\
    = \displaystyle{\int \frac{\sin (\Delta_+/2)}{\Delta_+/2}
    \varphi_{\scriptsize{\mbox{fp}}}(\Omega_1)\varphi_{\scriptsize{\mbox{fp}}}(\Omega_2)\xi^*(\Omega_3)
    \frac{\delta(\Omega_1+\Omega_2-\Omega_3-\Omega) }{1+\d\Omega_3^2}{\bf \mbox{d} \bar{\Omega}} +}\\ \displaystyle{ + \int \frac{\sin
    (\Delta_-/2)}{\Delta_-/2}
    2\varphi_{\scriptsize{\mbox{fp}}}(\Omega_1)\xi(\Omega_2)\varphi_{\scriptsize{\mbox{fp}}}^*(\Omega_3)
    \frac{\delta(\Omega_1-\Omega_2-\Omega_3+\Omega) }{1+\d\Omega_2^2}{\bf \mbox{d} \bar{\Omega}} }\\
    \Delta_{\pm} \equiv \Omega_1^2 \pm \Omega_2^2-\Omega_3^2 \mp \Omega^2\\
  \end{array}
\end{equation}

\begin{figure}[t!]
  \begin{picture}(370,207)(0,0)
   \put(15,10){\includegraphics[width=0.98\textwidth]{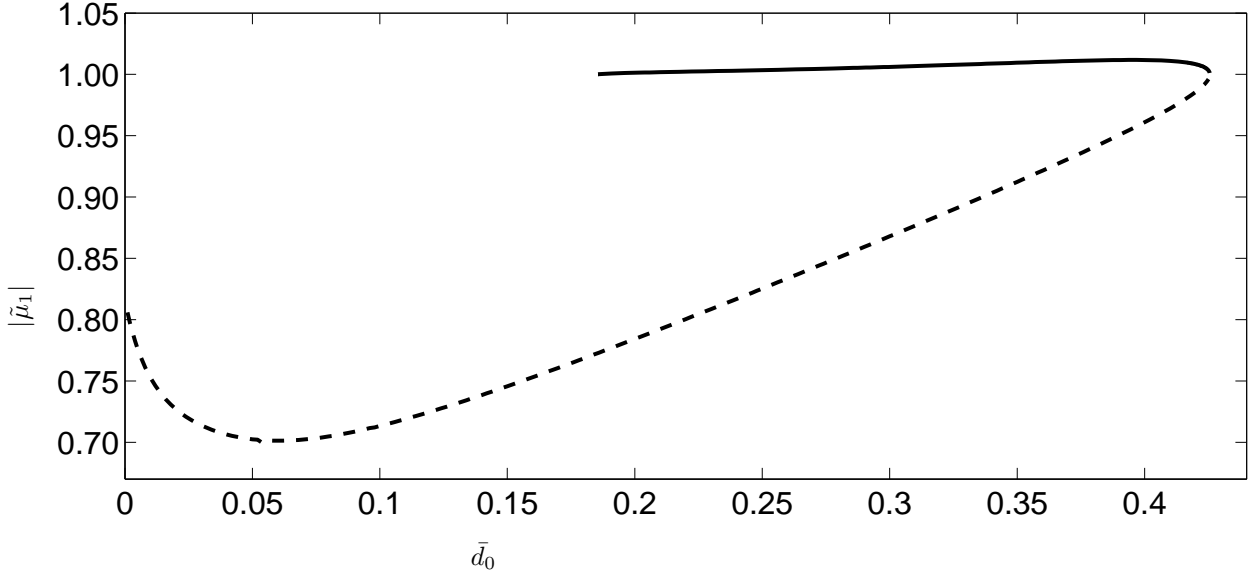}}
   \put(175,0){$\d$}
   \put(0, 90){\rotatebox{90}{$|\tilde{\mu}_1|$}}
  \end{picture}
  \caption{A plot of largest eigenvalue of $\hat{K}$ as a function of
    time $\d$ at the upper branch solutions (solid) and at the lower
    branch solutions (dashed).}
  \label{ul_max}
\end{figure}

\begin{figure}
  \begin{picture}(370,207)(0,0)
   \put(15,10){\includegraphics[width=0.98\textwidth]{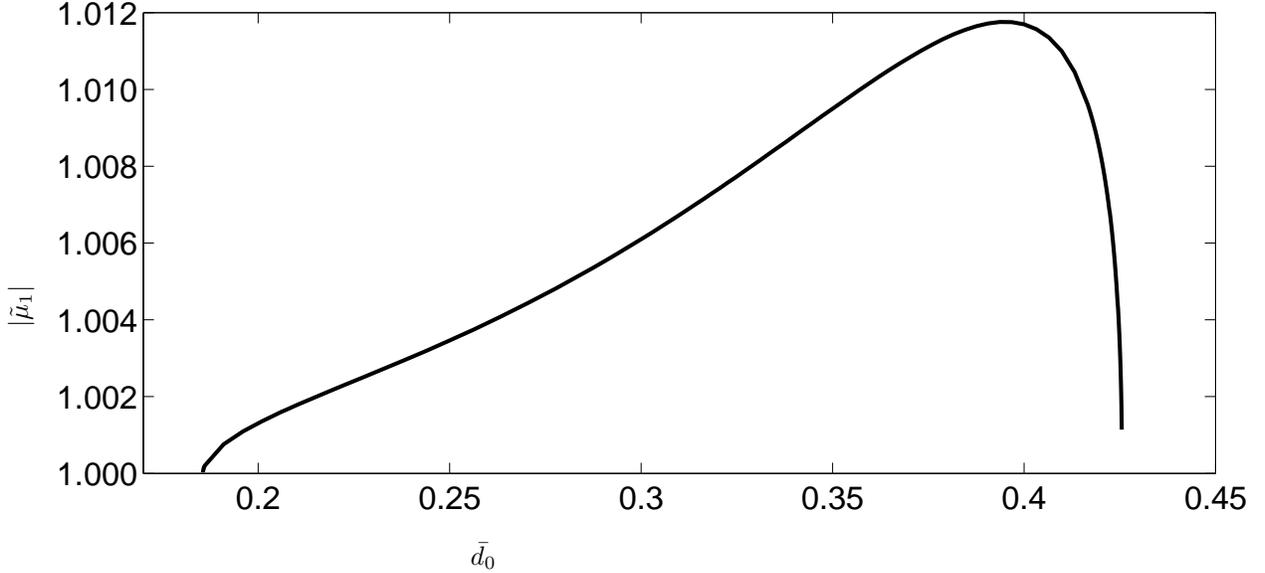}}
   \put(175,0){$\d$}
   \put(0, 90){\rotatebox{90}{$|\tilde{\mu}_1|$}}
  \end{picture}
  \caption{Plot of $|\tilde{\mu}_1|$ vs $\d$ for the upper branch only.
    The graph exhibits vertical behavior near $\d \approx 0.4256$ and
    near $\d \approx 0.1856$. In the first case this behavior is due to
    bifurcation.}
  \label{ub_max}
\end{figure}

Figure~\ref{ul_max} shows how the largest magnitude eigenvalue of the
operator $\hat{K}$ depends on $\d$. Here the dashed line corresponds to
the lower branch of solutions and the solid line corresponds to the
upper branch of solutions. Since for all values of $\d$ the largest
eigenvalue of $\hat{K}$ at the lower branch solutions is less than 1 the
map $N$ is contractive. Therefore as long as we are close enough to the
fixed point the iterations will converge. The situation is different for
the upper branch of solutions. As clearly shown on Figure~\ref{ub_max}
for all values of $\d$ the largest eigenvalue is larger than 1 and
therefore the map is not contractive.

\section{Construction of a contractive map}\label{polyn}

We construct a map that retains the same fixed points as $N$ and at the
same time a map that is contractive in the neighborhood of the upper
branch fixed points. Consider a polynomial $\mathbb{P}(x) =
\sum_{j=0}^{p} b_j x^{j}$ and use it to construct a new map P:
\begin{equation}
   \begin{array}{l}
    P(\varphi) \equiv \mathbb{P}(N(\varphi)) = \displaystyle{\sum_{j=0}^{p}b_j N^{j}(\varphi)}\\
    N^{0}(\varphi)=\varphi, ~~~ N^{j}=N(N(\ldots N(\varphi))) ~~~\mbox{$j$ times}
   \end{array}
\end{equation}
In the neighborhood of a fixed point this operator has the following
linearization
\begin{equation}
   \begin{array}{r}
     P(\varphi_{\scriptsize{\mbox{fp}}}+\psi) =\displaystyle{
     P\left(\varphi_{\scriptsize{\mbox{fp}}}+\sum_i a_i |\tilde{\phi}_i \rangle\right)= \sum_{j=0}^{p}b_j \left(\varphi_{\scriptsize{\mbox{fp}}}+K^{j}\left(\sum_i a_i |\tilde{\phi}_i \rangle \right) \right)}=\\
     = \displaystyle{\sum_{j=0}^{p}b_j\left(\varphi_{\scriptsize{\mbox{fp}}} + \sum_{i} a_{i} \tilde{\mu}_{i}^{j} |\tilde{\phi}_i \rangle \right) =\varphi_{\scriptsize{\mbox{fp}}}
     \sum_{j=0}^{p} b_j + \sum_i a_i \left(\sum_{j=0}^{p} b_j \tilde{\mu}_{i}^{j} \right)| \tilde{\phi}_i \rangle = }
   \end{array}
\end{equation}
We choose the values of $b_j$ so that the $\sum b_j = 1$. Due to this
condition the new map retains the same fixed point as $N$. Thus after
$n$ iterations we have
\begin{equation}
  P^n(\varphi_{\scriptsize{\mbox{fp}}}+\psi) =
    \varphi_{\scriptsize{\mbox{fp}}} + \sum_i a_i \left
    \{\mathbb{P}(\tilde{\mu}_i)\right \}^n |\tilde{\phi}_i \rangle
  \label{n_iterations}
\end{equation}
We choose the coefficients of the polynomial $b_i$ in such a manner that
the condition $\max\{|\mathbb{P}(\tilde{\mu}_i)|\}<1$ is satisfied.
According to equation~(\ref{n_iterations}) each mode decays
exponentially, and the smaller $\max \{|\mathbb{P}(\tilde{\mu}_i)| \}$
is, the faster the convergence.

Our approach is a modification of the technique considered
in~\cite{StepanovChernykh01}, where only first order polynomials were
needed, in our case we found that 5th order polynomials seemed to be
optimal. The coefficients of the polynomial could be optimized for the
faster convergence. This issue is discussed in the matrix Chebyshev
approximation problem~\cite{GreenLioydTref94, TrefethenBau97,
TohTref98}.

\begin{figure}
  \subfigure[]
   {
   \begin{picture}(370,98)(0,0)
   \put(15,10){\includegraphics[width=0.98\textwidth]{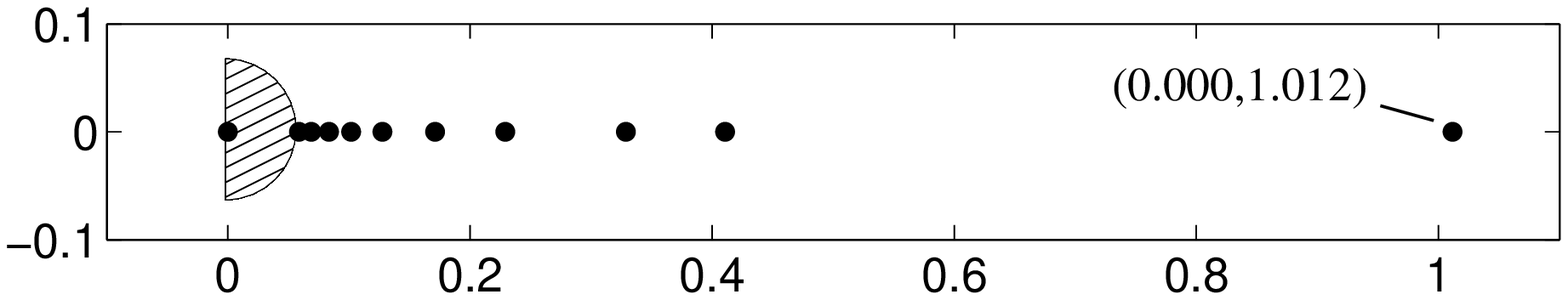}}
   \put(175,0){Re($\tilde{\mu}$)}
   \put(0,40){\rotatebox{90}{Im($\tilde{\mu}$)}}
   \end{picture}
   }
\subfigure[]
   {
   \begin{picture}(370,98)(0,0)
   \put(15,10){\includegraphics[width=0.96\textwidth]{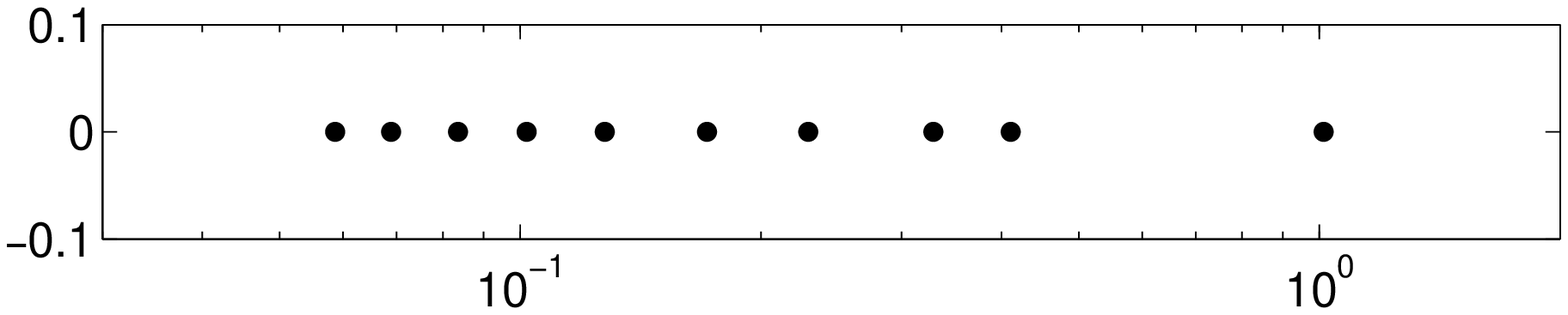}}
   \put(175,0){Re($\tilde{\mu}$)}
   \put(0,40){\rotatebox{90}{Im($\tilde{\mu}$)}}
   \end{picture}
   }
\caption{A plot of the imaginary part of eigenvalue versus the real part. Plotted first 10 eigenvalues of operator K at $\varphi_{\d =
0.4}^{\scriptsize{\mbox{upper}}}$. In subfigure (a) the shaded region defines the part of space where the rest of eigenvalues lay. In subfigure (b) the real part
is plotted on logarithmic scale.}
\label{spec_2_5}
\end{figure}
\begin{figure}
  \begin{picture}(370,205)(0,0)
   \put(15,10){\includegraphics[width=0.98\textwidth]{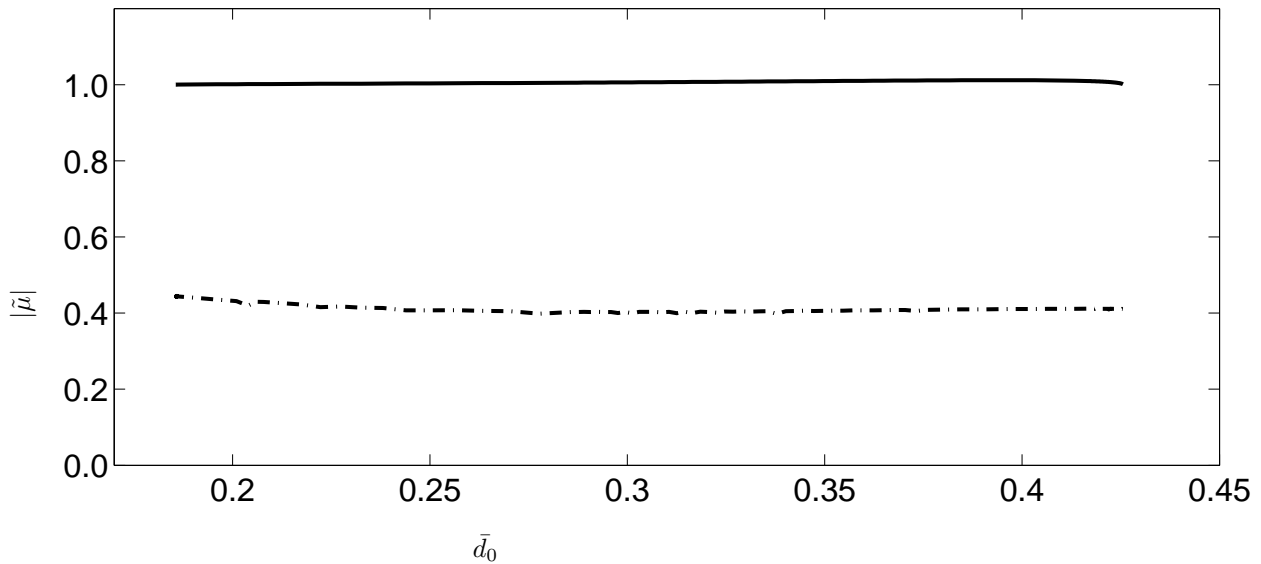}}
   \put(175,0){$\d$}
   \put(0,90){\rotatebox{90}{$|\tilde{\mu}|$}}
  \end{picture}
  \caption{A plot of $\tilde{\mu}_1$ (solid) and $\tilde{\mu}_2$
    (dash-dot) versus $\d$. This plot indicated a large gap between the
    first two eigenvalues for the upper branch solutions.}
  \label{ub_2max}
\end{figure}
\begin{figure}
  \begin{picture}(370,225)(0,0)
   \put(15,10){\includegraphics[width=0.98\textwidth]{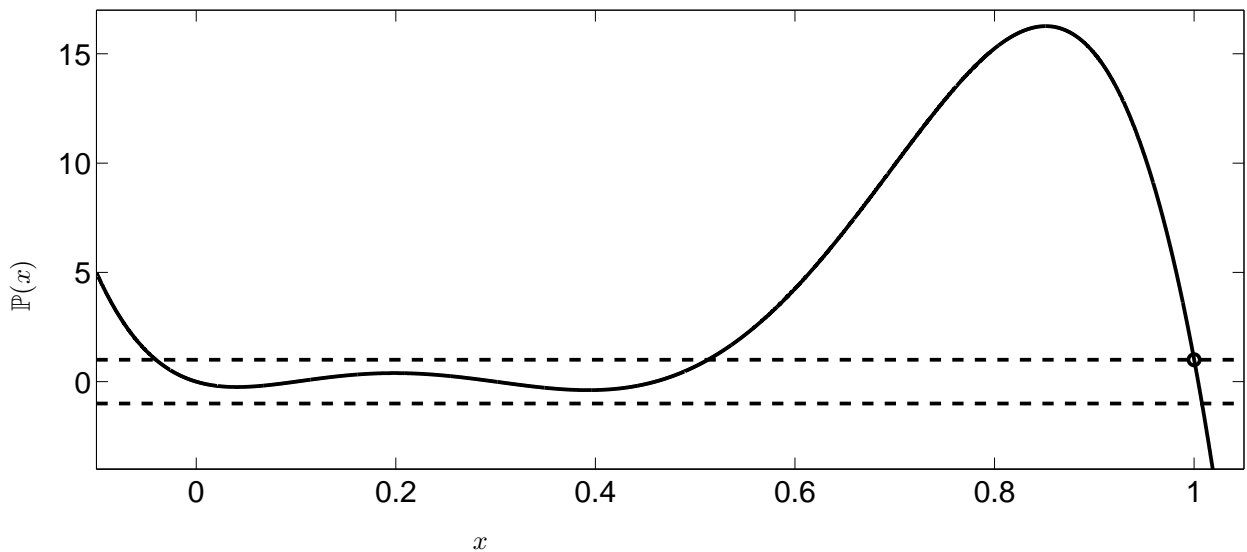}}
   \put(175,0){$x$}
   \put(0,90){\rotatebox{90}{$\mathbb{P}(x)$}}
  \end{picture}
  \caption{This is a plot of a polynomial suggested in
    equation~(\ref{polynomial}). Dashed lines indicate values of $1$ and
    $-1$, outside of these values eigenvalues of $\mathbb{P}(N)$ will be
    greater than 1, in which case the new map will not be contractive.}
  \label{poly}
\end{figure}

Figure~\ref{spec_2_5} shows the 10 largest eigenvalues of $K$ at the
fixed point $\varphi_{\d=0.4}^{\scriptsize{\mbox{upper}}}$ found
numerically using the method in equation~(\ref{eigenN}). This is a
typical picture for all upper branch fixed points. This computation
shows that the imaginary part of these eigenvalues is either very small
or zero. The real part of the first 10 eigenvalues lies between 0 and
1.012 and all other eigenvalues lie in the shaded area. There are two
key features of the spectrum that should be emphasized. First,
$\tilde{\mu}_1$ is extremely close to 1. As shown on figure~\ref{ub_max}
this feature is shared by all of the upper branch solutions. Second, as
demonstrated by figure~\ref{ub_2max} the following inequality takes
place: $|\tilde{\mu}_1-1|\ll|1-\tilde{\mu}_2|$. The decay of the first
mode is proportional to $\mathbb{P}(\tilde{\mu}_1)^n$ where $n$ is the
number of iterations performed, therefore, to make the convergence of
this mode as fast as possible we choose a polynomial with a derivative
at (1,1) near a value of $-1/(\tilde{\mu}_1-1)$. Due to the first
feature $\tilde{\mu}_1-1$ is very small, which implies fast growth of
the polynomial to the left of the point $(1,1)$. However, due to the
second feature of the spectrum such a behavior is not problematic, since
there is plenty of room for the polynomial to return close to zero.
Thus, consider the polynomial presented in the figure~\ref{poly} and
defined as
\begin{equation}
 \begin{array}{c}
  \mathbb{P}(x)=\displaystyle{\prod_{i=1}^{p} (x-r_i) {\bigg /}
  \prod_{i=1}^{p} (1-r_i)}\\ \mbox{and let } r_1 = 0.0, r_2 = 0.1, r_3 = 0.3, r_4 = 0.46, r_5 = 1.01
 \end{array} \label{polynomial}
\end{equation}
This is a fifth order polynomial with $|\mathbb{P}(\tilde{\mu}_i)|< 1$
and $\mathbb{P}(1)=1$. Following the earlier discussion we can see that
this polynomial does in fact has large slope at $(1,1)$, and yet there is
enough room for the values to drop back down as $x$ approaches
$\tilde{\mu}_2$. Also note that this is only a fifth degree
polynomial, and therefore one iteration of $\mathbb{P}(N)$ requires only
five iterations of $N$.

To catch the solution on the upper branch we have to be as close as
possible to the solution before we start our iterative process. Before
the computing the upper branch, the lower one were obtained.
Because of the bifurcation structure is was natural to expect that the
function $2\varphi_{\d_{\rm bf}} - \varphi_{\d}$ could serve as a good
starting seed for the upper branch solution at $\d$. Even with simple
iterations (without the polynomial deformation of the map in order to
make it contractive) it was clear that there exists another branch of
solutions, as for the first few iteration the residual (\ref{res}) was quickly decreasing (diverging in the long run.)

The behavior of $\tilde{\mu}_1$ illustrated in Figure~\ref{ub_max}
suggests that there is another bifurcation point with $\d_{\rm bf}
\approx 0.18$ and yet another branch of antisymmetric bisoliton
solutions. There is strong evidence of that from applying the shooting
$2\varphi_{\d_{\rm bf}} - \varphi_{\d}$ already from the computed upper
branch.

\section{Direct simulation of bisoliton propagation}

\begin{figure}
   \begin{picture}(370,210)(0,0)
     \put(15,10){\includegraphics[width=0.98\textwidth]{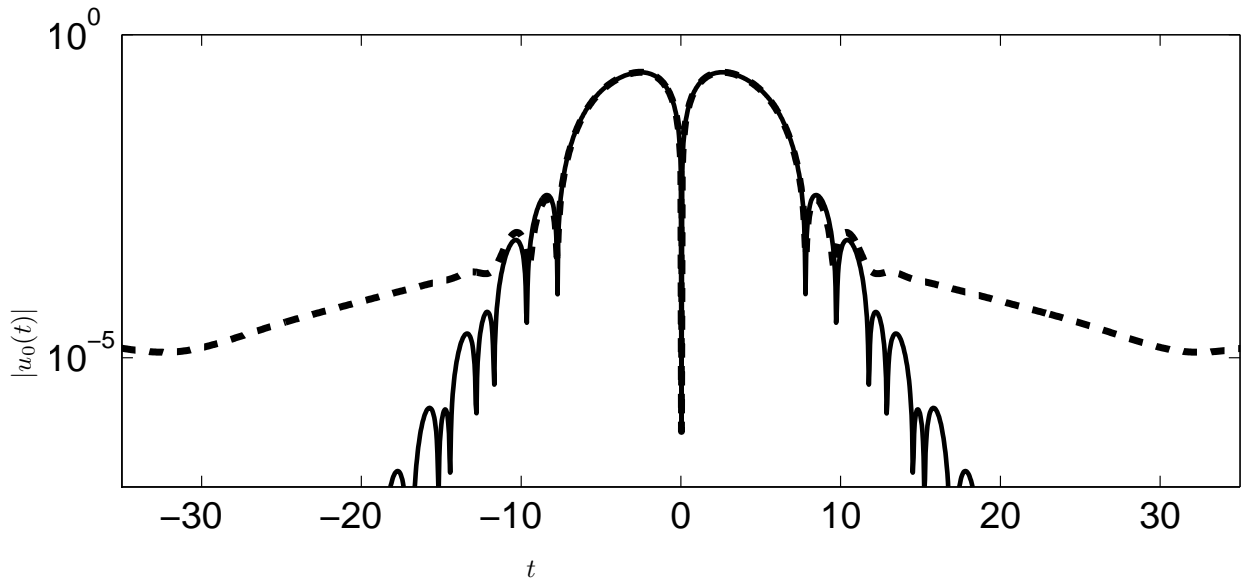}}
     \put(195,0){$t$}
     \put(0,75){\rotatebox{90}{$|u_0(t)|$}}
   \end{picture}
   \caption{Directly solving NLS equation with initial condition $u_0$
     (solid) from equation~(\ref{initial}) with lower branch solution
     $\varphi_{\d=0.214}$ $d_0=0.0125$, $d_1=1.25$, $L=1.2$ and $\gamma
     = 1$. The result of propagation $u_0$ over 3000 periods is
     represented by dashed line, showing three orders of magnitude
     agreement between the initial and final pulses. Here $P\approx
     0.0345$ so that
     $z_{\scriptsize{\mbox{rd}}}=45$ and
     $z_{\scriptsize{\mbox{nl}}}=29$.}
  \label{pulse}
\end{figure}

As a demonstration that the computed solutions of the averaged
equation~(\ref{Slow}) are solitary wave approximations of the
equation~(\ref{NLS}) we perform a direct simulation of~(\ref{NLS}) with
the following initial condition
\begin{equation}
  u_0(t) \equiv u(t,L/4) = 2 \pi \sqrt{P}F^{-1} \left. \left[ \varphi_{\d }
    (\Omega) \right] \right|_{\tilde{t} = t/\sqrt{L d_1/2}}
  \label{initial}
\end{equation}
(remember that $\d \equiv d_0 / ( \gamma P )(d_1 L/2)$). Here $F^{-1}$ is
the inverse Fourier transform. The pulse is launched at $z = L/4$
because as equation~(\ref{envelope_phase}) indicates at this point the
phase of the solution $\varphi_{\d}$ is constant. Notice that parameters
$d_0$, $d_1$, $L$ and $\gamma$ are determined by the physical system,
while $P$ is a free parameter whose value is limited from below due to
the upper bound on $\d = \d_{\scriptsize{\mbox{bf}}}$:
\begin{equation*}
  \frac{2d_0}{\d_{\scriptsize{\mbox{bf}}} d_1}\frac{1}{\gamma L} < P
    \ll \frac{1}{\gamma L}
\end{equation*}
The inequality on the right must be maintained in order to stay in the
regime where Gabitov-Turitsyn equation is valid. Figure~\ref{pulse}
shows the result of propagating $u_0$ over 3000 dispersion map periods.
This distance exceeds the distances needed for information transmission.
The solid line represents the initial pulse $u_0(t)$ of
equation~(\ref{initial}) launched at $z=L/4$. The pulse $u_0$ is
constructed using a fixed point $\varphi_{\d=0.4}$ of the map
$N_{\d=0.4}$. It is used as an initial condition for direct simulation
of equation~(\ref{NLS}) solved using a split-step method. The dashed
line shows an amplitude profile, plotted on a logarithmic scale, of the
pulse after the distance $3000L$, showing excellent agreement between
the initial and final pulses and confirming that $u_0(t)$ is indeed a
good approximation to a solitary wave solution of equation~(\ref{NLS}).

\section{Conclusion}

In this paper we have presented a method to calculate solitary wave
solutions of the nonlinear \Schrod equation with periodic dispersion and
weak nonlinearity. In case when the effect of local dispersion is much
stronger than residual dispersion and nonlinearity the \Schrod equation
is reduced to Gabitov-Turitsyn equation, an integro-differential
equation. This equation is satisfied by the first order term in the
asymptotic expansion of a solution to the Schr$\ddot{\mbox{o}}$dinger
equation with periodic dispersion. Finding solitary wave solutions is
then reduced to finding a solution of an integral equation by finding
fixed points of a derived map.

We investigated bound pairs of solitary waves, bisolitons, and showed
that system has at least two branches of nontrivial solutions. On one of
these branches, our map is not contractive. We introduce a method to
polynomially modify the map, obtaining new map that retains the original
fixed points, while being contractive. Effective use of this method
resulted in a previously unknown branch of bisoliton solutions. In order
to apply this method the spectrum of the map was studied. We carefully
studied the parameter space where such solutions are possible, and found
two places where solutions appear to bifurcate.

The method of polynomial deformation as applied to our problem can be
further improved. The choice of the polynomials can be adapted as the
knowledge about the spectrum improves in the process of iterations.
Also, a better precision on the eigenvalues may allow one to further
decrease the order of the polynomial, thus expediting the iteration
procedure. The approach taken by this method is not limited to the
antisymmetric solutions of our equation. Higher order solutions can be
found after the appropriate modification of the map. Thus we have used this method to study symmetric bisolitons, with the results to be published elsewhere.

Authors would like to thank Ildar Gabitov and Linn Mollenauer for posing
the problem, and Robert Indik and Sasha Korotkevich for the useful
discussions during the work on this problem. This work supported in part
by the State of Arizona grant TRIF (Proposition 301), NSFgrant
EMSW21-VIGRE \# 0602173, and by the Department of Energy at Los Alamos
National Laboratory under contracts DE-AC52-06NA25396 and the DOE Office
of Science Advanced Computing Research (ASCR) program in Applied
Mathematical Sciences.

\section*{Appendix A}
\label{appendix}

Here we present another way of reducing the equation~(\ref{NLS}) to a
dimensionless form. In the $\omega$-space we introduce a slowly varying
amplitude $q(z, \omega)$ as
\begin{eqnarray}
  && u(z, \omega) = q(z, \omega) \, \exp \left( - {\rm i} \omega^2
    \int_{L/4}^z {\rm d}z' \Big( d(z') - \langle d \rangle \Big) \right) .
    \nonumber
\end{eqnarray}
Here the angular brackets denote averaging over $z$. The slow (over many
periods of the dispersion map) dynamics of $q$ is described by
Gabitov-Turitsyn equation~\cite{GT_96}
\begin{eqnarray}
  && {\rm i} q_{z}(\omega) - \langle d \rangle \omega^2 q(\omega) +
    \gamma \int \frac{{\rm d}\omega_1 {\rm d}\omega_2 {\rm
    d}\omega_3}{(2\pi)^2} \delta(\omega_1 + \omega_2 - \omega_3 - \omega)
    S(\delta) q(\omega_1) q(\omega_2) q^*(\omega_3) = 0
    , \nonumber \\
  && \delta = \omega_1^2 + \omega_2^2 - \omega_3^2 - \omega^2 , \nonumber
    \quad
  S(\delta) = \left\langle \exp \left( - {\rm i} \delta \int_{L/4}^z
    {\rm d}z' \Big( d(z') - \langle d \rangle \Big) \right) \right\rangle
    . \nonumber
\end{eqnarray}
For the dispersion map $d(z) = d_0 + d_1$ if $0 < z < L / 2$, and $d(z)
= d_0 - d_1$ if $L / 2 < z < L$ we get $S(\delta) = \sin(\delta L d_1 /
4) / (\delta L d_1 / 4)$. We choose $\sqrt{L d_1 / 2}$ to be a unit of
time and introduce the dimensionless frequency $\Omega = \sqrt{L d_1 /
2} \omega$. Then $S(\Delta) = \sin(\Delta / 2) / (\Delta / 2)$, where
$\Delta = \Omega_1^2 + \Omega_2^2 - \Omega_3^2 - \Omega^2$.

We set the residual dispersion to be equal to $1 / 2$ by choosing the
unit of propagation distance: $Z = (4 d_0 / L d_1) z$. The nonlinearity
coefficient $\gamma$ can be eliminated by choosing the unit of the pulse
amplitude $u$ (or $q$): $q(z, \omega) = \sqrt{2 d_0 / \gamma} \, Q(Z,
\Omega)$.

The dimensionless form of the Gabitov-Turitsyn equation becomes
\begin{eqnarray}
  && {\rm i} Q_{Z}(\Omega) - {\textstyle \frac12} \Omega^2 Q(\Omega) +
    \int \frac{{\rm d}\Omega_1 {\rm d}\Omega_2 {\rm
    d}\Omega_3}{(2\pi)^2} \delta(\Omega_1 + \Omega_2 - \Omega_3 -
    \Omega) \frac{\sin(\Delta/2)}{\Delta/2} Q(\Omega_1) Q(\Omega_2)
    Q^*(\Omega_3) = 0.
  \label{dimless_eq}
\end{eqnarray}
The stationary shape solutions have the form $Q(Z, \Omega) = {\rm
e}^{{\rm i} \Lambda Z} \varphi(\Omega)$ --- there is a family of
solutions parameterized by wave number $\Lambda$.

In the main point of this addendum is to present another way to bring
the equation~(\ref{NLS}) to the dimensionless form. Here the final step
in reducing the number of parameters is done by choosing the wave number
of the inverse solution as the unit of propagation distance. This way
the only remaining parameter is dimensionless residual dispersion
$\bar{d}_0$ that is related to the wave number of the solution
of~(\ref{dimless_eq}) $\Lambda$ as $\bar{d}_0 = 1 / 2\Lambda$. The
advantage of this approach is that we now study the family of solutions
of one equation (\ref{dimless_eq}) instead of considering a family of
equations (parametrized by $\bar{d}_0$), where for each of these
equations a solution is found with a certain wave number.


\end{document}